# Superconductivity at 5K in $NdO_{0.5}F_{0.5}BiS_2$

Rajveer Jha, Anuj Kumar, Shiva Kumar Singh and V. P. S. Awana

*Quantum Phenomena and Application Division, National Physical Laboratory* (*CSIR*)

*Dr. K.S. Krishnan Road, New Delhi-110012, India*

We report appearance of superconductivity at 5K in $NdO_{0.5}F_{0.5}BiS_2$ and supplement the discovery [1] of the same in layered sulfide based ZrCuSiAs type compounds. The bulk polycrystalline compound is synthesized by conventional solid state route via vacuum encapsulation technique. Detailed structural analysis showed that the studied compound is crystallized in tetragonal *P4/nmm* space group with lattice parameters $a$ = 3.9911(3)Å, $c$ = 13.3830(2)Å. Bulk superconductivity is established in $NdO_{0.5}F_{0.5}BiS_2$ at 5K by both transport and magnetic measurements. Electrical transport measurements showed superconducting $T_c$ onset at 5.2K and $T_c$ ($\rho$=0) at 4.7K. Under applied magnetic field both $T_c$ onset and $T_c$ ($\rho$ = 0) decrease to lower temperatures and an upper critical field [$H_{c2}(0)$] of above 23kOe is estimated. Both *AC* and *DC* magnetic susceptibility measurements showed bulk superconductivity below 5K. Isothermal magnetization (*MH*) exhibited typical type-II behavior with lower critical field ($H_{c1}$) of around 15Oe. Specific heat [$C_p(T)$] is investigated in the temperature range of 1.9-50K in zero external magnetic field. A Schottky-type anomaly is observed at low temperature below 7K. This low temperature Schottky can be attributed to the change in the entropy of the system.

*Corresponding Author
Dr. V. P. S. Awana, Senior Scientist
E-mail: awana@mail.npindia.org
Ph. +91-11-45609357, Fax-+91-11-45609310
Homepage www.freewebs.com/vpsawana/

Very recent reports on appearance of $BiS_2$ based layered superconductivity in $Bi_4O_4S_3$ [2-5] and $REO_{0.5}F_{0.5}BiS_2$ (RE = La, Nd, Pr, Ce and Yb) [1, 6-11] is of tremendous interest. This is because these compounds are layered in structure and similar to that as popular high $T_c$ cuprates [12] and Fe-pnictides [13]. In fact preliminary electronic structure calculations show that even the carrier doping mechanism is similar to that as for cuprates and Fe-pnictides [8, 14-16]. It seems the role of $Cu-O_2$ planes of cuprates and FeAs of pnictides is being played by $BiS_2$ in the newly discovered compounds. Presuming this, one wonders if the relatively low temperature (below 10K) superconductivity of these new systems could meet the same fate as of cuprates and Fe-pnictides and result into higher $T_c$. This may open a fresh avenue for an altogether new series of $BiS_2$ based exotic superconductors. At, the situation is that with normal pressure synthesis the $BiS_2$ superconductivity is found at above 4.5K in $Bi_4O_4S_3$ [2-5] and at 2.7K [6-7] and 5.5K [1] in $LaO_{0.5}F_{0.5}BiS_2$ and $NdO_{0.5}F_{0.5}BiS_2$ respectively. Under applied pressure the 2.7K superconductivity of $LaO_{0.5}F_{0.5}BiS_2$ increases to above 10K [8]. Also, with high pressure high temperature (*HPHT*) solid state synthesis route stable superconductivity of above 10K could be established in $LaO_{0.5}F_{0.5}BiS_2$ [6]. Worth mentioning is the fact that possibility of foreign impurity material has also been suggested for the exotic layered $BiS_2$ based superconductivity in $Bi_4O_4S_3$ [17]. However, the possibility of impurity (Bi) driven superconductivity is discarded in ref. [3] as it (impurity phase Bi) is present in ordinary non-superconducting rhombohedral phase in the matrix. On the same footing, jump at $T_c$ in $C_p$ measurement for $LaO_{0.5}F_{0.5}BiS_2$ [9] and $Bi_4O_4S_3$ [20] confirmed the bulk superconductivity of these $BiS_2$ based compounds.

Interestingly, the $NdO_{0.5}F_{0.5}BiS_2$ compound which exhibits superconductivity above 5K without *HPHT* treatment (normal pressure synthesis) [1] yet awaits its confirmation from independent groups. Band structural calculations suggest that ground state of these newly discovered $La/NdOBiS_2$ (1112) is not superconducting but rather semiconducting [14] similar to that as for high $T_c$ cuprates compounds [12]. Experimentally as well the ground state $La/NdOBiS_2$ is found non-superconducting [1, 6-8]. In fact lot more need to be done to explore fully the ground state of these newly discovered Bismuth oxy sulfides.

In this article we report the synthesis and superconductivity of nearly phase pure $NdO_{0.5}F_{0.5}BiS_2$ compound and substantiate the results of reference [1]. The as synthesized $NdO_{0.5}F_{0.5}BiS_2$ compound is nearly single phase in nature and bulk superconducting below 5K as established by both electrical transport and magnetization measurements. In heat capacity measurements a Schottky anomaly is observed below 7K, which suppresses the jump at $T_c$ due to superconducting transition.

Bulk polycrystalline NdO$_{0.5}$F$_{0.5}$BiS$_2$ is synthesized by standard solid state reaction route via vacuum encapsulation. High purity Nd, Bi, S, NdF$_3$, and Nd$_2$O$_3$ are weighed in stoichiometric ratio and ground thoroughly in a glove box under high purity argon atmosphere. The mixed powders were subsequently palletized and vacuum-sealed (10$^{-3}$ Torr) in a quartz tube. Sealed quartz ampoule is placed in tube furnace and heat treated at 800$^0$C for 12h with the typical heating rate of 2$^o$C/min., and subsequently cooled down slowly over a span of six hours to room temperature. This process was repeated twice. X-ray diffraction (*XRD*) was performed at room temperature in the scattering angular (*2θ*) range of 10$^o$-80$^o$ in equal *2θ* step of 0.02$^o$ using *Rigaku Diffractometer* with *Cu K$_α$* ($\lambda$ = 1.54Å). Rietveld analysis was performed using the standard *FullProf* program. Detailed electrical transport and magnetization measurements were performed on Physical Property Measurements System (*PPMS*-14T, *Quantum Design*) as a function of both temperature and applied magnetic field. Heat capacity of the studied sample over a temperature range 1.9-50K, was also performed on *PPMS*-14T.

Figure 1 (a) depicts the room temperature observed and Rietveld fitted *XRD* pattern of studied NdO$_{0.5}$F$_{0.5}$BiS$_2$ sample. The compound is crystallized in ZrCuSiAs type tetragonal structure in space group *P4/nmm*. Rietveld refinement of *XRD* pattern is carried out using ZrCuSiAs structure and Wyckoff positions. Refined structural parameters (lattice parameters, atomic coordinates and site occupancy), are shown in the Table I. The representative unit cell of the compound in *P4/nmm* space group crystallization is shown in Figure 1 (b). The layered structure includes Nd$_2$O$_2$ (rare earth oxide layer) and BiS$_2$ (fluorite type) layers. Various atoms with their respective positions are indicated in the inset part of Figure 1. Bismuth (Bi), Neodymium (Nd), and Sulfur (S1 and S2) atoms occupy the *2c* (0.25, 0.25, *z*) site. On the other hand O/F atoms are at *2a* (0.75, 0.25, 0) site. In NdO$_{0.5}$F$_{0.5}$BiS$_2$, mobile carriers are doped from NdO/F redox layer to superconducting BiS$_2$ layer. It is worth mentioning here that the phase purity of the sample is achieved after several heating schedules and a slight change in synthesis temperature gives rise to large amount of Nd$_2$O$_3$ impurity and non superconducting behavior.

Both *DC* and *AC* magnetization with temperature results are shown in Fig. 2. As it can be seen, the compound exhibits diamagnetic transition at below 5K. There is a positive back ground as well seen in the normal state (above 5K) magnetic data. This is presumably due to the paramagnetic moment of magnetic ion Nd being present in superconducting NdO$_{0.5}$F$_{0.5}$BiS$_2$ sample. As far as the volume fraction of superconductivity is concerned the same is considerably less than another BiS$_2$ based superconductor, i.e., Bi$_4$O$_4$S$_3$ with shielding fraction of around 95% [2-3]. The volume fraction of superconductivity in reported La/Nd-1112 (La/NdOBiS$_2$) is yet weak [1, 6-8] and better refinement of the material quality is yet warranted. However, an interesting analogy comes here in mind, when BiS$_2$ based 1112 are compared with FeAs based 1111 (REFeAsO), the $T_c$ of former increases from 2.7K [6-8] to above 5K [1] when La is changed with Nd, in later case the same goes from say 27K [13] to above 50K [18]. It means the structural pressure plays an important role in case of layered BiS$_2$ based 1112 compounds, similar to that as for FeAs based 1111 pnictides. The lower inset of figure shows the isothermal magnetization (*MH*) of the compound at 2.1K. NdO$_{0.5}$F$_{0.5}$BiS$_2$ clearly exhibits the type-II behavior with its lower critical field ($H_{c1}$) at around 15Oe. The normal state magnetic behavior of the NdO$_{0.5}$F$_{0.5}$BiS$_2$ sample is shown in upper inset Fig. 2. The magnetic susceptibility of the compound increases with decrease in temperature. This is due to paramagnetic moment of Nd in studied NdO$_{0.5}$F$_{0.5}$BiS$_2$ sample.

Fig.3 depicts the resistivity versus temperature (*ρ*-T) plots for studied NdO$_{0.5}$F$_{0.5}$BiS$_2$ sample with and without applied magnetic field of up to 20kOe in temperature range of 2-7K Inset of the same shows the resistivity in expanded temperature range 2-300K. The compound is semiconducting/non-metallic down to 6K. The superconducting onset in terms of sharp decrease in resistivity is seen at 5.2K with $T_c$ ($ρ$ = 0) at 4.7K without any applied external field. The normal state conduction and the superconducting transition temperature are in agreement with the reported results in ref. 1. With applied magnetic field both $T_c$(onset) and $T_c$($ρ$ = 0) are decreased to lower temperatures.

Fig. 4 shows the d*ρ*/d*T* plot of the sample. Single characteristic peak is seen in d*ρ*/d*T* plots at various fields although with increased broadening under applied field. This is suggestive of single superconducting transition and better grains coupling in this system. The broadening of d*ρ*/d*T* peak under applied field suggests that superconducting onset is relatively affected less than the $T_c$ ($ρ$ =0) state. In any case, with the application of magnetic field both the onset and offset $T_c$ shift towards lower temperature. Inset of Fig. 4 shows the variation of upper critical field $H_{c2}(T)$ and irreversibility field $H_{irr}(T)$. $H_{c2}(T)$ and $H_{irr}(T)$ are calculated through the intersection point between normal state resistivity extrapolation and superconducting transition line respectively. The slope $dH_{c2}/dT|_{Tc}$ is estimated to be ~ 7.5*k*Oe/K that implies *WHH* (Werthamer-Helfand-Hohenberg) $H_{c2}(0)$ (= - 0.69 T$_c$ $dH_{c2}/dT|_{Tc}$) value of 23*k*Oe, which is slightly than the one (19*k*Oe) reported for LaO$_{0.5}$F$_{0.5}$BiS$_2$ of [7].

Fig. 5 represents the heat capacity measurement [$C_p$(T)] with temperature of NdO$_{0.5}$F$_{0.5}$BiS$_2$. Hump at $T_c$ related to superconducting transition is not observed in heat capacity measurements and rather a Schottky-type anomaly is observed at below ~7K. This may be due to some incipient magnetic ordering, which is also appearing in magnetization measurement. This ordering may be dominating over superconducting transition of $C_p$. Similar type of Schottky anomaly is also observed for $CeO_{0.5}F_{0.5}BiS_2$ compound [9]. More rigorous experimental and theoretical studies are warranted to invoke these compounds. However, the jump in $C_p$ measurement is observed for $LaO_{0.5}F_{0.5}BiS_2$ [9] and for $Bi_4O_4S_3$ [19] confirming the bulk superconductivity of these $BiS_2$ based compounds, though samples have small electronic specific heat coefficient. In this context it is also noticeable that very weak jump at $T_c$ in $C_p$ measurement had been reported for well established pnictide superconductors [20] and in layered nitrides [21]. The inset (a) of Fig. 5 represents $C_p/T$ vs T plot and inset (b) shows the $C_p/T$ vs $T^2$ plot. By the fitting of $C_p$ data we estimated the value of Sommerfeld constant (γ) and the coefficient of the lattice contribution (β), as 26.23 *mJ/mole-$K^2$* and 1.428 *mJ/mole-$K^4$* respectively.

In conclusion we have synthesized a near single phase $NdO_{0.5}F_{0.5}BiS_2$ compound and bulk superconducting below 5K is established by both electrical transport and magnetization measurements. The compound is a type II superconductor with lower [$H_{c1}$(2K)] and upper [$H_{c2}$(0K)] critical fields of around 15Oe and 23kOe respectively. Schottky-type anomaly is observed in $C_p$ measurement below ~7K due to some magnetic ordering, which may also be suppressing superconducting jump at $T_c$.

**Table 1** Atomic coordinates, Wyckoff positions, and site occupancy for studied $NdO_{0.5}F_{0.5}BiS_2$.

| Atom | x | y | z | site | Occupancy |
|------|---|---|---|------|-----------|
| Nd | 0.2500 | 0.2500 | 0.098(4) | 2c | 1 |
| Bi | 0.2500 | 0.2500 | 0.625(5) | 2c | 1 |
| S1 | 0.2500 | 0.2500 | 0.365(1) | 2c | 1 |
| S2 | 0.2500 | 0.2500 | 0.831(6) | 2c | 1 |
| O | 0.7500 | 0.2500 | 0.000 | 2a | 0.5 |
| F | 0.7500 | 0.2500 | 0.000 | 2a | 0.5 |

**Figure Captions**

Figure 1 (a): Observed (*open circles*) and calculated (*solid lines*) *XRD* pattern of $NdO_{0.5}F_{0.5}BiS_2$ compound at room temperature.

Figure 1 (b): Schematic unit cell of $NdO_{0.5}F_{0.5}BiS_2$ compound.

Figure 2: AC magnetic susceptibility and *DC* magnetization (both *ZFC* and *FC)* plots for $NdO_{0.5}F_{0.5}BiS_2$, lower inset shows isothermal *MH* curve of the sample at 2.1K, marking its lower critical field ($H_{c1}$). Upper inset shows the normal state paramagnetic behavior of the compound.

Figure 3: Resistivity versus temperature ($\rho$-T) plots for $NdO_{0.5}F_{0.5}BiS_2$ sample with and without applied magnetic field of up to 20kOe. Inset shows the resistivity in expanded temperature range 2-300K.

Figure 4: Variation the d$\rho$/dT with temperature of $NdO_{0.5}F_{0.5}BiS_2$ sample. Inset of Fig. 4 shows the variation of upper critical field $H_{c2}(T)$ and irreversibility field $H_{c2}(T)$ and $H_{irr}(T)$.

Figure 5: Specific heat $C_p$ versus temperature T for $NdO_{0.5}F_{0.5}BiS_2$. Lower inset shows the plot of $C_p/T$ versus $T^2$ and fitting. The plot for $C_p/T$ versus T is shown in the upper inset of the figure.

**Figure 1** (a)

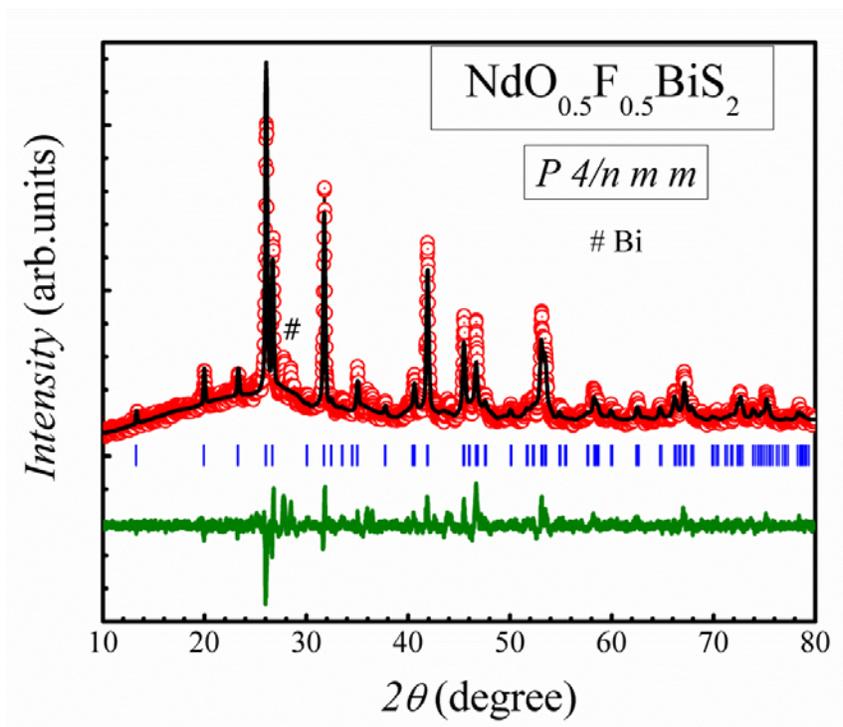

**Figure 1 (b)**

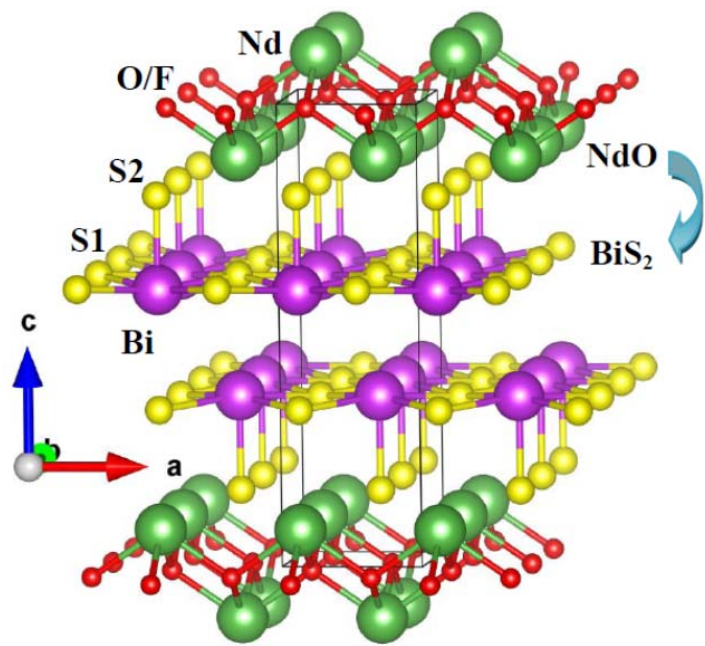

**Figure 2**

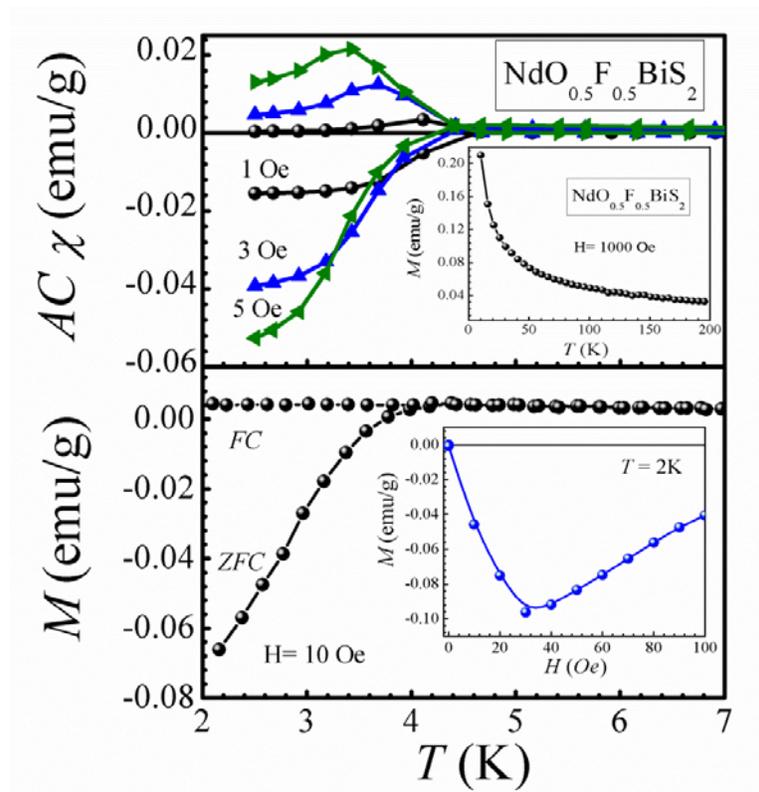

**Figure 3**

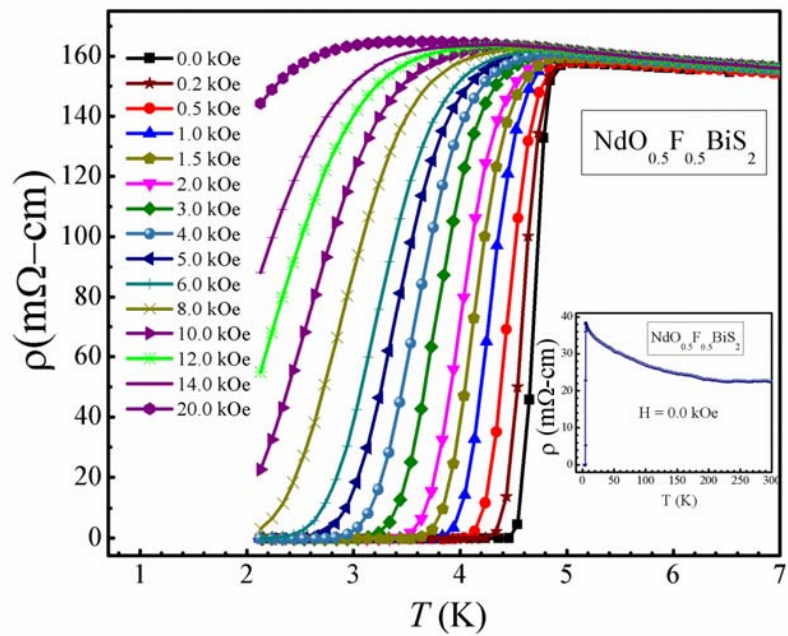

**Figure 4**

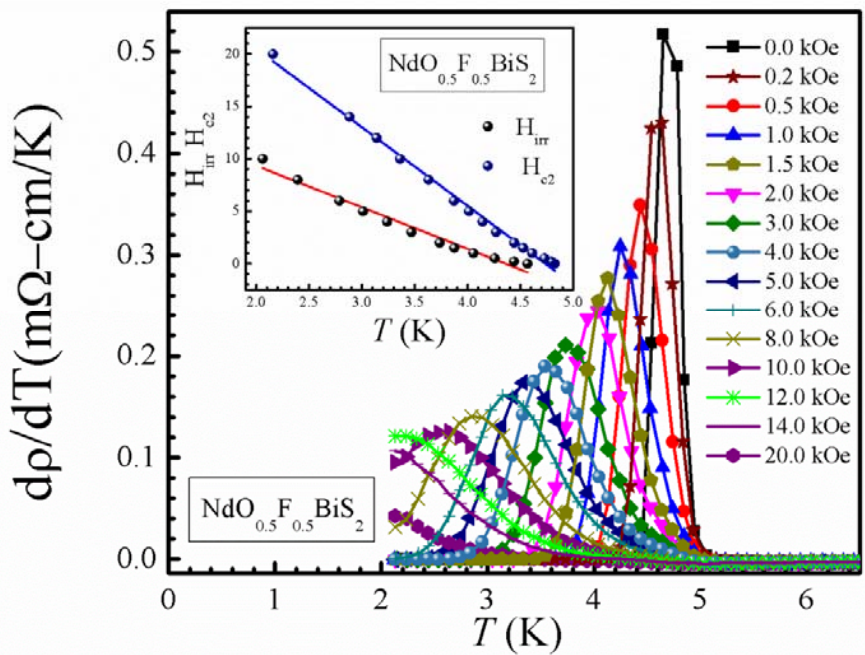

**Figure 5**

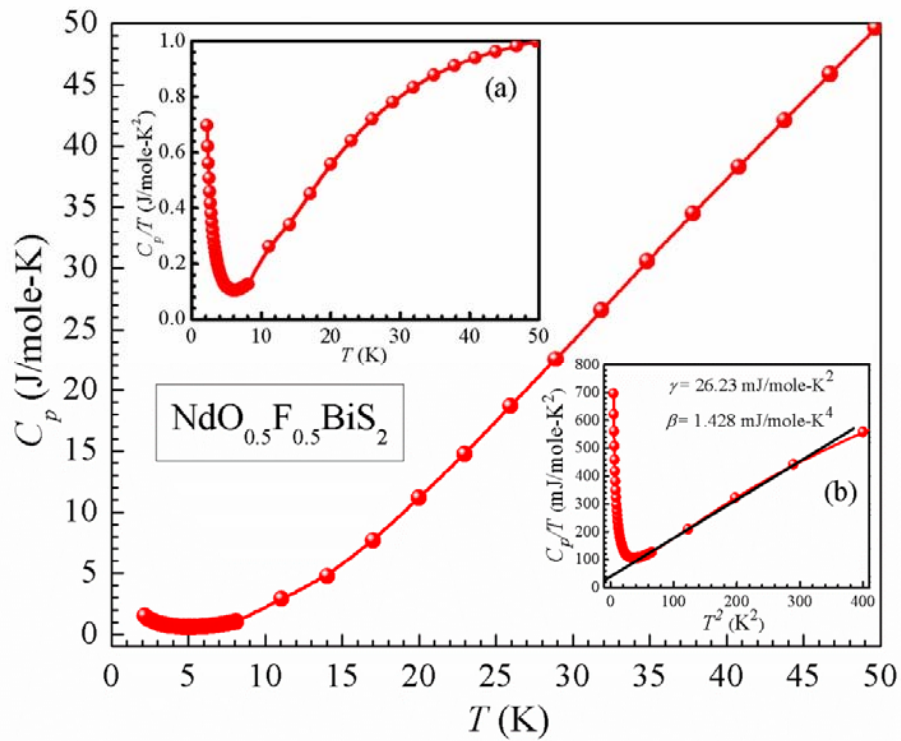